\documentclass[a4paper,12pt]{article}
\usepackage{amsmath}
\usepackage[pdftex]{graphicx}
\usepackage{subcaption}
\usepackage{chicago}
\title{Spontaneous cooperation for public goods}

\author{Jeroen Bruggeman, Rudolf Sprik, Rick Quax\thanks{Jeroen Bruggeman: Department of Sociology, University of Amsterdam, Nieuwe Achtergracht 166, 1018 WV Amsterdam, Netherlands. Corresponding author: \texttt{j.p.bruggeman@uva.nl}. Rudolf Sprik: van der Waals-Zeeman Institute, Institute of Physics, University of Amsterdam, Science Park 904, 1098 XH Amsterdam, Netherlands. Rick Quax: Informatics Institute, Faculty of Science, University of Amsterdam, Science Park 904,  1098 XH Amsterdam, Netherlands. }}
\date{\today}
\begin{document}  
\maketitle

% >>>> Variation S_0

% {\bf Keywords:} Collective action, inter-group conflict, uncertainty, conformity, Ising model

% voor J. Math. Soc.: haal critical mass naar voren en maak bounded rationality en noise uitgangspunten.  Jackson e.a. (2018) sync & arousal als empirisch ondersteuning

% Consensus on S_0 (and delta) by social influence model that we have anyway; behandel als common frame of reference en als focal point (Schelling)? Test variatie - zie Hale 2013 Ann Rev Pol Sci

\begin{abstract} Cooperation for public goods poses a dilemma, where individuals are tempted to free ride on others' contributions. Classic solutions involve monitoring,  reputation maintenance and costly incentives, but there are important collective actions based on simple and cheap cues only, for example unplanned protests and revolts. This can be explained by an Ising model with the assumption that individuals in uncertain situations tend to conform to the local majority in their network. Among initial defectors, noise such as rumors or opponents' provocations causes some of them to cooperate accidentally. At a critical level of noise, these cooperators trigger a cascade of cooperation. We find an analytic relationship between the phase transition and the asymmetry of the Ising model, which in turn reflects the asymmetry of cooperation and defection. This study thereby shows that in principle, the dilemma of cooperation can be solved by nothing more than a portion of random noise, without rational decision making. \end{abstract}

People may want to realize or preserve public goods, for example democracy and clean air, but because contributors are disadvantaged in the face of free riders, there is a dilemma \cite{olson65,hardin68,ostrom09,gavrilets15}. Solutions typically require efforts of the participants to monitor one another \cite{rustagi10} and spread information (gossip) \cite{nowak05} through their network reliably that establishes reputations \cite{panchanathan04}, upon which some of them have to deliver individual rewards or (threats of) punishments \cite{fehr03}, under pro-social norms to preclude arbitrariness. These provisions are not always (sufficiently) available, though, whereas in certain situations, participants still manage to self-organize cooperation, even without leaders. 
% All this comes on top of the costs of contributing to the public good. 
% In some situations, participants self-organize into cooperation without costly social mechanisms or reliable information. 
Cases in point are impromptu help at disasters, non-organized revolts against political regimes \cite{lohmann94,tilly02,tufekci17} and spontaneous street fights between groups of young men. % Also several animal species manage to cooperate without costly mechanisms, for example buffalo herd bulls who chase away prowling lions \cite{estes91}. 

These examples have in common a high uncertainty of outcomes, and unknown benefits and costs. Rational decision making is therefore not feasible. Participants who identify with their group or its goal \cite{stekelenburg13}, and thereby feel group solidarity \cite{durkheim12}, use the heuristic of \textit{conformism} to the majority of their network neighbors \shortcite{wu14}, which can be based on no more than visual information.  Human ancestors lived in groups for millions of years \cite{shultz11} and in all likelihood, solidarity and conformism are both cultural and genetic \cite{boyd18}. On an evolutionary time scale, conformism must have been beneficial on average when future benefits and costs were unknown \cite{vandenberg18}. To explain cooperation under conformism, we use an Ising model \shortcite{weidlich71,galam82,stauffer08,dorogovtsev08,castellano09}. This model recovers the critical mass that makes cooperation self-reinforcing but without the rationality assumptions of critical mass theory \cite{marwell93}.

\section*{Model}
A group $g$ of individuals who share an interest in a public good is modeled as a network with weighted and usually asymmetric ties $A_{ij}$ denoting $i$ paying attention to $j$. There is no assumption that $i$ and $j$ know each other before they meet at the site where collective action might take place. Consistent with other models of social influence \cite{friedkin11}, the adjacency matrix is row-normalized, yielding cell values $ a_{ij} = A_{ij} / \sum_j A_{ij}$, hence $ \sum_j a_{ij}$ = 1. 

Individuals have two behavioral options, defect ($D$) and cooperate ($C$), $ C > D > 0$, and all defect at the start.  The average degree of cooperation among $n$ individuals is described by an order parameter $ M = 1/n \sum_{i = 1}^{n} S_i $, where the behavioral variable $S_i$ can take the value $ S_i=C $ or $ S_i=-D $, for example $S=\{1,-1/2\}$. 
Everybody gets an equal share of the public good but cooperators incur a cost. A widely used definition of payoffs for cooperators $\Pi^g_C$ and defectors $\Pi^g_D$ in a group $g$ is the following \shortcite{perc17},
\begin{eqnarray}
\Pi^g_C &=&   r (N_C + 1)/n - 1, \nonumber \\
\Pi^g_D &=&   r  N_C / n,
\label{eq:payGame}
\end{eqnarray}
with $r > 1$ an enhancement, or synergy, factor of cooperation, 
and $N_C$ the number of cooperators when the focal player decides. In our case, 
\begin{eqnarray}
P^g_C &=&   r(M + C/n) - C, \nonumber \\
P^g_D &=&   r(M - D/n),
\label{eq:payGame2}
\end{eqnarray} 
 % (In the symmetric Ising model, with $C = D$, defining $r=1/(2C)$ assures that $rM_{max} - 1 = rM_{min}$.) 
which is identical to Eq.~\ref{eq:payGame} in standard game theory when $C = 1$ and $D = 0$.
Our payoffs can be negative but that does not matter because they are used only comparatively. Other, for example non-linear, payoff functions \cite{marwell93} may also be used. The key point, however, is that under high uncertainty,
participants do not maximize their payoff (directly) but align with others instead, thereby forming a collective lever that increases their payoff while avoiding exploitation. Behavior and network ties are expressed in the conventional, but here asymmetric, Ising model
\begin{equation}
H = - \sum_{i \neq j}^{n} a_{ij} S_i S_j.
\label{eq:ising1}
\end{equation}  
Solving the model boils down to minimizing $H$, where $H/n$ can be interpreted as average dissatisfaction. Minimizing can be done computationally with a Metropolis algorithm \cite{barrat08}, where individuals decide sequentially as in many network models, or through a mean field analysis, elaborated below. 

% Dilemma's of cooperation are usually analyzed by game theory in terms of payoffs \cite{perc17}. A group of defectors at a baseline payoff level, usually zero, can get higher payoffs (in our case lower $H$) if all, or most, individuals cooperate, but because nobody wants to be exploited by free riders while being tempted to free ride themselves, it is a non-trivial challenge to reach the cooperative state. Our Ising model has these two states in common with game theory, and expresses the difficulty to reach the cooperative state in terms of a (free energy) barrier, illustrated qualitatively in Fig.~\ref{fig:coop}(a).  After analyzing the model's behavior, we will elaborate the model's relationship to payoffs. 

High-uncertainty situations to which the model applies are characterized by  \textit{turmoil}, $T$, or temperature in the original model. It causes arousal, measurable as heart rates \shortcite{konvalinka11}, and produces noise \cite{lewenstein92} in individuals' information about the situation, which in turn becomes partly false, ambiguous, exaggerated or objectively irrelevant. Turmoil and its noise may consist of rumors, fire, provocations and violence. Some social movements produce turmoil by themselves, for instance an increasingly frequent posting of online messages \shortcite{johnson16}.
Arousal and noise entail ``trembling hands" \cite{dion88} as game theorists say, which means a chance that some individuals accidentally change their behavior. The model is to show that few accidental cooperators entail a cascade of cooperation. 
An example of turmoil and its ramifications is the self-immolation of a street vendor in December 2010, which, in the given circumstances, set off the Tunisian revolution. Other examples are the revolts in East Germany \cite{lohmann94} and Romania in 1989 and in Egypt and Syria in 2011 \cite{hussain13}, where protesters were agitated by rumors about the events in neighboring countries. Autocratic rulers try to prevent revolts by suppressing turmoil, for instance by tightening media control. 

Noise is different from a stable bias, for instance the ideology of an autocratic regime, which entails revolts against it less often than a weakened regime, or stumbling opponents in street fights. Opponents' weakness gives off noisy signals that they might be overcome, which readily entail collective actions against them \cite{skocpol79,goldstone01,collins08}. Whereas responses to noise are typically spontaneous, collective responses to stable signals, biased or not, tend to be mounted by organized groups with norms, incentives and all that \cite{tilly02,tufekci17,goldstone01}.  Combinations of signal and noise also occur, of course, which can result in, for example, an organized peaceful demonstration to suddenly turn violent. Our focus is on spontaneous cooperation.

\begin{figure}
\begin{center}
\includegraphics[width=0.8\textwidth]{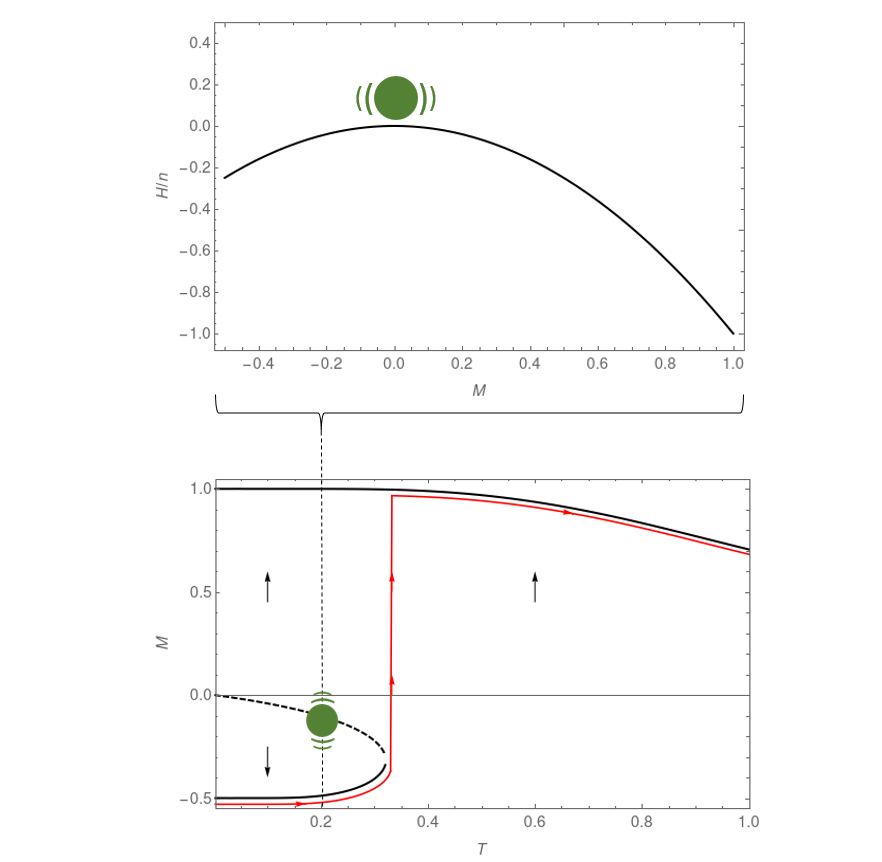} 
\end{center}
\caption{Cooperation for public goods. (a) Mean dissatisfaction $H/n$ and level of cooperation $M$. When all defect,  $H/n$ is at a local minimum, on the left, but to proceed to the global minimum where all cooperate, on the right, participants are hindered by a hill. (b) Mean field analysis with $ S = \{1,-1/2\} $ shows below $T_c$ one stable state with mostly cooperators, at the top, and another stable state in finite time with mostly defectors, at the bottom. A metastable state in between, indicated by the dotted line, corresponds to the hill top in Fig.~1(a). Above $T_c$ only one state remains, where with increasing $T$, cooperators are joined by increasing numbers of defectors.} 
\label{fig:coop}
\end{figure}

\begin{figure}
\centering
\begin{subfigure}[b]{2.6in}
\includegraphics[width=0.99\textwidth]{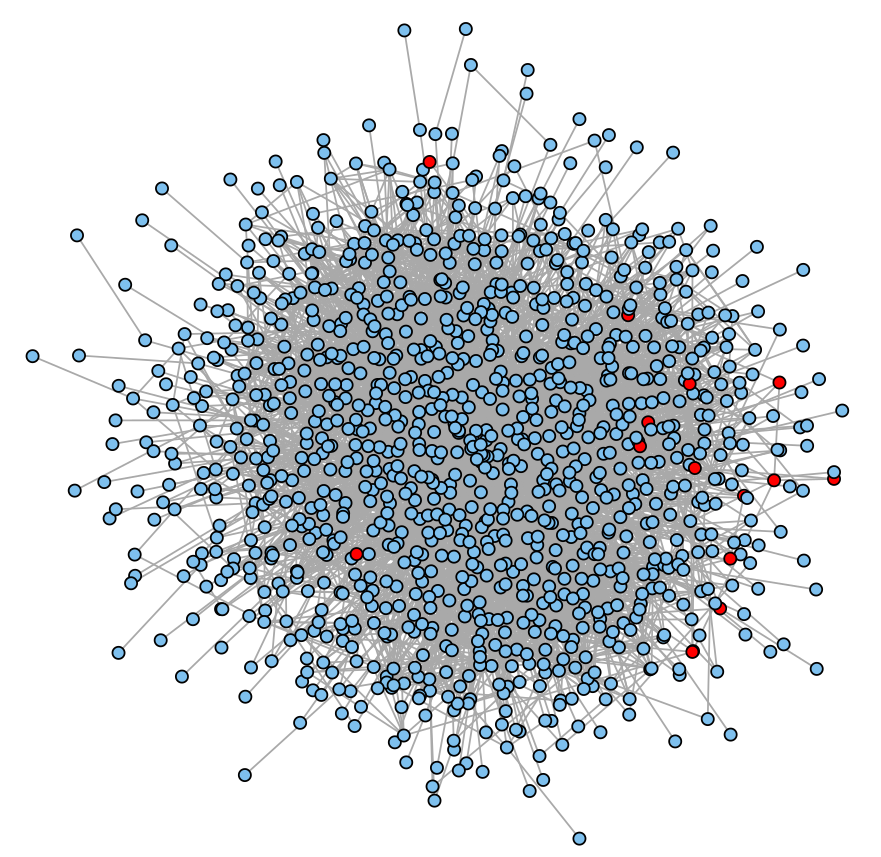}
\caption{}
\end{subfigure}
\begin{subfigure}[b]{2.6in}
\includegraphics[width=0.99\textwidth]{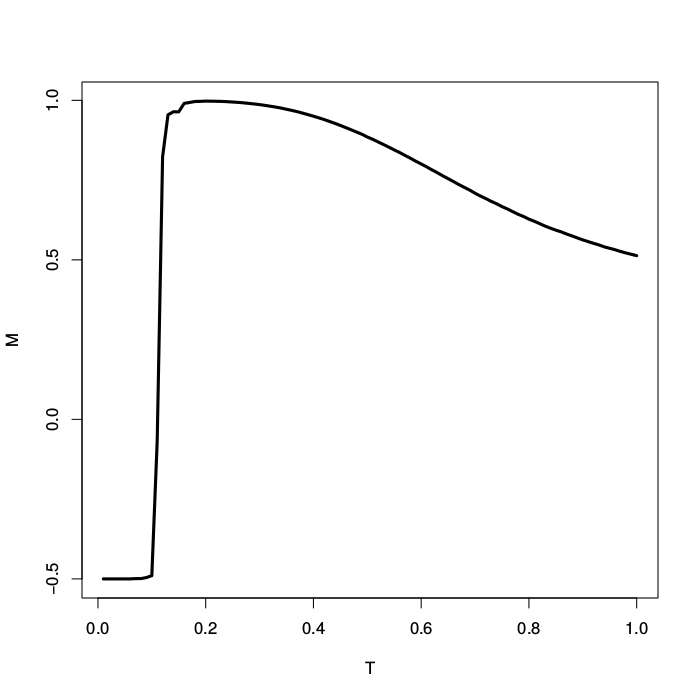}
\caption{}
\end{subfigure}
\caption{Numerical simulation. (a) a social network with $n=1133$ and clustering $C=0.254$ % \cite{guimera03} 
at a near-critical turmoil level of $T = 0.101$; $ S = \{1,-1/2\} $. Some nodes start cooperating (red) whereas most still defect (blue). (b) $T_c$ is smaller than in the mean field approximation, but the overall pattern is qualitatively the same (compare to Fig.~\ref{fig:coop} b).}
\label{fig:numerical}
\end{figure} 

\section*{Results}
\begin{figure}
\centering
\begin{subfigure}[b]{2.6in}
\includegraphics[width=0.99\textwidth]{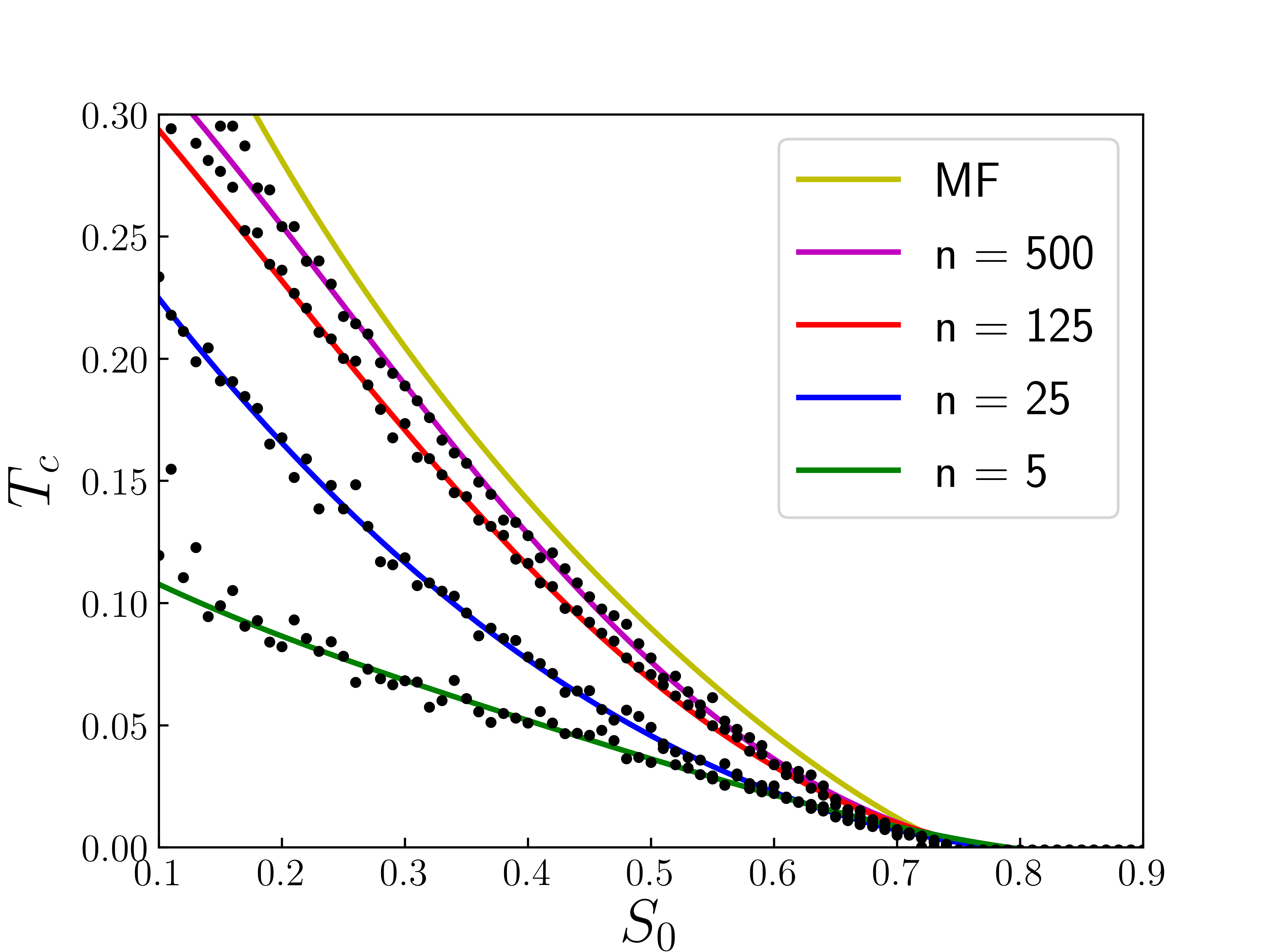} 
\caption{}
\end{subfigure}
\begin{subfigure}[b]{2.6in}
\includegraphics[width=0.99\textwidth]{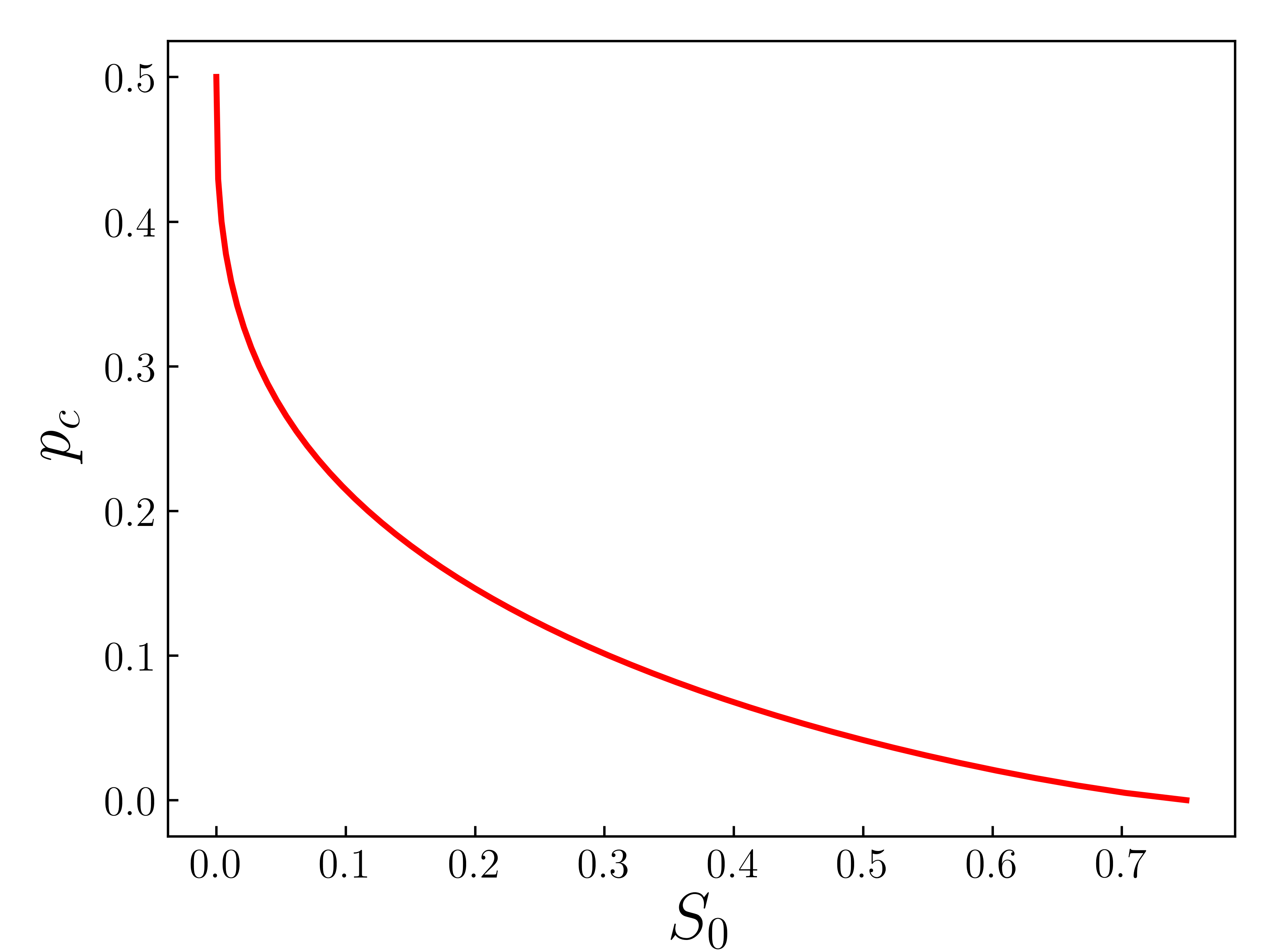} 
\caption{}
\end{subfigure}
\caption{Consequences of shifting $S_0$ with the mean field approach (MF), keeping $\Delta = 0.75$. (a) With increasing $S_0$, less agitation is necessary to turn defectors into cooperators. For comparison, numerical simulations on several random networks with density = 0.8 are shown as well.
(b) The proportion of defectors $p_c$ at $T_c$ decreases with increasing $S_0$.} 
\label{fig:result}
\end{figure}

Our general result is that within finite time and at low turmoil, cooperation does not get off the ground, but it does emerge at a critical level $T_{c}$. This pattern is illustrated by the red line in Fig.~\ref{fig:coop}(b) along the direction of the arrows. The figure was obtained with a mean field approach, but numerical simulations with the Metropolis algorithm show up the same pattern, with lower $T_c$ for small networks (Fig~\ref{fig:result}a). The model thus shows that at a critical level of turmoil, few accidental cooperators can trigger a cascade of cooperation.    
If $T$ keeps increasing way beyond $T_c$, cooperators co-exist with increasing numbers of defectors, until the two behaviors become equally frequent.  If in actuality cooperation then collapses completely is an issue for further study. Otherwise cooperation ends when the public good is achieved, the participants run out of steam, or others intervene. 

Alternatively, if participants get to understand an enduring situation, their uncertainty will reduce and they may start acting strategically, which requires pro-social norms to prevent. If the participants then develop such norms prescribing rewards and punishments, these norms can be easily modeled as field(s) by adding term(s) $ - h \sum S_i$ to the Hamiltonian (Eq.\ref{eq:ising1}). Consequently, cooperation emerges without phase transition. The actual maintenance of these norms, however, will entail additional costs over and above the contributions, whereas spontaneous cooperation is relatively cheap.    

\subsection*{Comparison with the symmetric Ising model}
Rewriting the asymmetric Ising model in a symmetric form enables a direct comparison with results in the literature for symmetric models, and a generalization to arbitrary values of $S = \{C,-D\}$. A model with asymmetric values can be reformulated as a symmetric model with an offset, or bias, $S_0 = (C-D)/2$ and an increment $\Delta = (C+D)/2$ by the mapping
\begin{equation}
S = \{C, -D\} \rightarrow \{ S_0 + \Delta, S_0 - \Delta \}
\end{equation}
Accordingly, the values chosen in Fig.~\ref{fig:coop}(b), $S = \{1,-1/2\}$, imply $\Delta = 0.75$ and $S_0 = 0.25$. For given $\Delta$, increasing $S_0$ means increasing interest in the public good, or, in line with the literature on protests, increasing grievances \cite{stekelenburg13}.

Substitution of $S_0$ and $\hat{S}_i$ chosen from $\{ \Delta, -\Delta \}$ in $H$ yields
\begin{equation}
H = - \sum_{ij} a_{ij} (S_0 + \hat{S}_i)(S_0 + \hat{S}_j). 
\end{equation}
Expanding $H$ in orders of $S_0$ yields
\begin{equation}
- \sum_{ij} a_{ij} \hat{S}_{i} \hat{S}_{j} - S_0 \left( \sum_{ij} a_{ij} \hat{S}_{i} + \sum_{ij} a_{ij} \hat{S}_{j} \right) - S_0^2 \sum_{ij} a_{ij}.
\end{equation}
The first term in the expansion  $H_{sym} = - \sum_{ij} a_{ij} \hat{S}_{i} \hat{S}_{j}$ is a symmetric model with the same adjacency matrix as the original asymmetric model. The second term $H_{loc} =  - S_0 ( \sum_{ij} a_{ij} \hat{S}_{i} + \sum_{ij} a_{ij} \hat{S}_{j} )$ is proportional to $S_0$ and can be interpreted as a local field that modifies $H_{sym}$. The contribution of this local field can be expressed in terms of row and column sums of $a_{ij}$ as
\begin{equation}
H_{loc} = -S_0 \sum_{i}  \left(  \sum_{j} a_{ij} + \sum_{j} a_{ji}  \right )   \hat{S}_{i}.
\end{equation}
For row-normalized adjacency matrices, with $\sum_{j} a_{ij} = 1$ for all rows $i$, $H_{loc}$ becomes
\begin{equation}
H^{row}_{loc} = -S_0 \sum_{i}  \left(  \sum_{j} a_{ij} \right)  \hat{S}_{i} - S_0 \sum_{i} \hat{S}_{i},
\end{equation}
where the first term is a local field varying for each $ \hat{S}_{i} $, and the second term is a homogeneous external field independent of $a_{ij}$. The third term in the expansion of $H$ is independent of the values of $\hat{S}$ and is a constant depending on $a_{ij}$ only. Hence it does not play a role in the minimization of $H$. 
For a connected network with row-normalization, the last expression can be further simplified to
\begin{equation}
H^{row}_{loc} = -2 S_0 \sum_{i} \hat{S}_{i}.
\end{equation}
The asymmetry in $S$ is then equivalent to a symmetric system with an external field $2 S_0$. 

\subsection*{Mean field analysis}
The expected value of $M$ as function of $T$ can be obtained by assuming that the network is very large and by abstracting away from its topology; in the language of thermodynamics, by approximating the interaction energy by the energy of one spin (here, behavior) in the mean field of its neighbors \cite{barrat08}, $M = \langle S \rangle$. 
The value of $M$ can now be expressed in closed form in terms of the probabilities given by the exponential of the Hamiltonian energy and $T$ as
\begin{equation}
M = \frac{   (S_{0}-\Delta) e^{ \frac{(S_{0} - \Delta)M}{T}  } 
           + (S_{0}+\Delta) e^{ \frac{(S_{0} + \Delta)M}{T}  }     }
         {                  e^{ \frac{(S_{0} - \Delta)M}{T}  }
                          + e^{ \frac{(S_{0} + \Delta)M}{T}  }     }.
\label{Seqn1}
\end{equation}
This reduces to an implicit relation
\begin{equation}
\frac{M}{\Delta} = \frac{S_0}{\Delta} + \tanh\left(\frac{ \Delta^2 }{T} \frac{M}{\Delta} \right),
\label{Seqn2}
\end{equation}
where only dimensionless ratios of $M$, $S_{0}$ and $T$ with $\Delta$ remain in the expression. The mean degree $ \langle k \rangle$, defined for binary ties, does not occur in it because the adjacency matrix is row-normalized and the mean weighted outdegree $\langle k_w \rangle = 1$.

By analyzing the intersection of the line defined by $M/\Delta - S_0/\Delta$ and the tanh term on the right hand side of Eq.~\ref{Seqn2}, the possible values for $M$ at a given $T$ can be found. For $T>T_c$ there is one stable high $T$ solution and for $T<T_c$ there is one stable solution of (nearly) full cooperation, another solution that is stable in finite time with (nearly) full defection, and one unstable solution. At $T=T_c$ the two stable solutions merge and the intersecting line coincides with the tangent line touching the tanh function; see Fig.~\ref{fig:coop}(b). At that point a closed relation for $T_c$ in terms of $S_0$ and $\Delta$ can be found, 
\begin{equation}
\frac{ S_{0} }{ \Delta } = 
\frac{ 
 \sqrt{ \frac{\gamma -1}{\gamma+1} } (\gamma+1) \gamma - \cosh^{-1} ( \gamma )}
{\gamma^2}, 
\label{Seq3}
\end{equation}
where $\gamma = \sqrt { \frac{\Delta^{2}}{T_c} }. $ Eq.~\ref{Seq3} is used in Fig.~\ref{fig:result}(a). It shows that if $S_0$ increases while keeping $\Delta$ constant, less agitation is required to motivate defectors to cooperate.  When $ \Delta $ decreases to $ \Delta = S_0$, defection loses its appeal. The figure also shows that numerical simulations yield very similar results for large networks but diverge for small ones. This also holds true for $T_c$ in Fig~\ref{fig:coop}, which is lower for smaller networks (not shown).

From the mean field approximation follows the proportion of defectors $p_c$ and cooperators $1 - p_c$ at given $S_0$ and pertaining $T_c$, after a time long enough for the system to settle down. The proportion of cooperators $1 - p_c$ at $T_c$ is the critical mass \cite{marwell93}, and can be inferred from the value of $M_{c}$ at the phase transition,
\begin{equation}
M_{c} = p_{c}(S_0 - \Delta) +  (1-p_{c})(S_{0} + \Delta).
\label{Mc1}
\end{equation}
The mean field analysis of $T_c$ yields
\begin{equation}
\frac{M_{c}}{\Delta} = \frac{\rm{ - cosh}^{-1}(\gamma)}{\gamma^{2}}.
\label{Mc2}
\end{equation}
Note that the $\rm{cosh}^{-1}$ function in Eq.~\ref{Mc2} only yields a result when $\gamma > 1$, and sets a limit to $T_{c}$ for given $\Delta$. For the choice $\Delta=0.75$, the maximum value of $T_{c} = \Delta^2 = 0.565$.  Solving for $p_{c}$ yields
\begin{equation}
p_{c} = \frac{1}{2} - \frac{S_{0}}{2 \Delta} + \frac{1}{2} \frac{\rm{cosh}^{-1}(\gamma)}{\gamma^{2}},
\end{equation}
used for Fig.~\ref{fig:result}(b). It shows that the proportion of defectors $p_c$ at $T_c$ decreases with increasing $S_0$. In contrast to critical mass theory, however, the Ising model has no assumptions about initiative takers or leaders who win over the rest, rational decision making \cite{marwell93}, or learning that would require fairly stable feedback \cite{macy91}.

\section*{Discussion and Conclusion}
We have shown that under high uncertainty, the dilemma of collective action can be solved by nothing more than a portion of random noise. The asymmetric Ising model does not require any knowledge or accurate expectations of the participants, and only depends on conformism, which can be empirically observed in synchronous motion, gestures or shouting \cite{mcneill95,jones13}. In particular, it has no assumptions about actors' rationality, in contrast to critical mass theory, whereas it supports that theory's key findings of the critical mass and the tipping point. Simulations add to our mean field result that 
turmoil-driven cooperation is most likely in small groups, where cooperation starts at relatively low levels of noise. 

Shortly before we finished this manuscript, two other papers appeared where an Ising model was used to solve this dilemma \cite{adami18,sarkar18}, but their symmetric model requires complex quantum physics to define payoffs, in contrast to our simple definition. % Moreover, actual payoffs also depend on opponents' actions and changing circumstances that are rarely known in advance.  

In another paper, the effects of different networks and of multiple $S_0$ values on the tipping point are investigated \cite{brugspin2}. Along with more empirical testing, perhaps also on other species, a future direction might be to explicitly model noisy information transmission on the group's network \cite{quax13}, too.

% The asymmetric Ising model provides an explanation of spontaneous cooperation in uncertain situations with few and general assumptions on psychology and information. It may therefore apply to multiple species.

% \ldots when the usual provisions, in particular reliable information, institutions and selective incentives, are insufficient or non-available. Instead of rational decision makers, individuals under uncertainty are assumed to be conformists, responding to their immediate social environment with a margin of error determined by random noise. They may have expectations of payoffs that have consequences for the critical mass and the tipping point (Fig.~\ref{fig:result}).  

\subsection*{Acknowledgements}
Thank you to Jos\'{e} A. Cuesta, Raheel Dhattiwala, Don Weenink, and Alex van Venrooij for comments.

\subsection*{Author contributions}
JB wrote the paper and made the asymmetric Ising model. RS related the asymmetric model to the symmetric one. RQ did the mean field analysis. 

\small
\bibliographystyle{chicago}

% \bibliography{evo}

\end{document}